\documentclass[a4paper]{article}
\usepackage{fullpage}
\usepackage{amsmath,amsbsy,amssymb}
\newcommand{\Rho}{\mathrm{P}}

\title{On a direct approach to quasideterminant solutions of a noncommutative KP equation}
\author{ C. R. Gilson and J. J. C. Nimmo \\
Department of Mathematics, \\
University of Glasgow \\
Glasgow G12 8QW, UK}
\date{}

\begin{document}

\maketitle

\begin{abstract}
A noncommutative version of the KP equation and two families of
its solutions expressed as quasideterminants are discussed. The
origin of these solutions is explained by means of Darboux and
binary Darboux transformations. Additionally, it is shown that
these solutions may also be verified directly. This approach is
reminiscent of the wronskian technique used for the Hirota
bilinear form of the regular, commutative KP equation but, in the
noncommutative case, no bilinearising transformation is available.
\end{abstract}

\section{Introduction}\label{Introduction}
There has been recent interest in a noncommutative version of the
Kadomtsev-Petviashvili equation (ncKP)
\cite{HT1,P,S,WW1,WW2,WW3,H,K}
\begin{equation}\label{nckp}
(v_t+v_{xxx}+3 v_x v_x)_x+3  v_{yy}-
 3 [v_x,v_y] =0.
\end{equation}
This equation can be obtained via the compatibility of the same
Lax pair \eqref{lax1}--\eqref{lax2} as is used in the commutative
case, but it is not assumed that $v$ and its derivatives commute.
In the case that variables do commute, we may differentiate
\eqref{nckp} with respect to $x$ and set $v_x=u$ to obtain the
well-known (commuting) KP equation
\begin{equation}\label{kp}
(u_t+u_{xxx}+6uu_x)_x+3u_{yy}=0.
\end{equation}

In most of the recent work on ncKP the noncommutativity arises
because of a quantization of the phase space in which independent
variables do not commute and the (commutative) product of real- or
complex-valued functions of these are replaced by the associative
but noncommutative Moyal star product.  This approach is useful
from the point of view of interpreting solutions as they can be
expressed in terms of standard functions. It is however
conceptually quite difficult because of the noncommutativity of
the independent variables.

In this present paper we will not specify the nature of the
noncommutativity and the results we present are valid not only in
the star product case but also for, for example, the matrix or
quarternion versions of the KP equation. This is in the spirit of
the work by Etingof, Gelfand and Retakh \cite{EGR1997} in which
solutions of the ncKP equation were found in terms of
quasideterminants \cite{GR,GGRL} using a noncommutative version of
Gelfand-Dickey theory. Very recently, Hamanaka \cite{H2006} has
used this form of solution to obtain the soliton solutions of ncKP
in the Moyal product case.

We will consider two types of quasideterminant solutions of ncKP.
One is equivalent to those found in \cite{EGR1997} which we will
call \emph{quasiwronskians}. We will also consider a new type of
quasideterminant solution which we term a \emph{quasigrammian}.
These two types of solution are each constructed by iterating
Darboux transformations; the quasiwronskians using a standard
Darboux transformation and the quasigrammians using the related
binary Darboux transformation. The connection between Darboux
transformations for a matrix Schr\"odinger equation and
quasideterminants was also investigated in \cite{GV}.

We will then show that, in fact, these solutions can be verified
by direct substitution. This sort of direct approach is very
widely studied in the commutative case (see \cite{FN} for the
first results for the KP case, and \cite{Hirota book} for a
discussion of many other examples). In these cases one first makes
a change of dependent variable which converts the nonlinear
equation to Hirota bilinear form. For the KP equation one writes
$u=2(\log\tau)_{xx}$ and then \eqref{kp} is converted to Hirota
form, a homogeneously quadratic differential equation in $\tau$. A
solution $\tau$ in the form of a determinant may be verified by
recognising the Hirota form as a determinantal identity such as a
Pl\"ucker relation or Jacobi identity.

In contrast, the central role of the $\tau$-function (a
determinant) in the commutative case is taken by the
quasideterminant in the noncommutative case. In this case there is
no bilinearising change of variables since $v$ is expressed
directly as a quasideterminant. Also, remarkably given what
happens in the commutative case, no use is made of special
identities in verifying the quasideterminant solutions of ncKP.
Some remarks on the reason for this are given later.
Paradoxically, direct verification of solutions in the
noncommutative case is in a number of respects easier than in the
commutative case. However, this state of affairs seems to be
particular to ncKP. In other examples we have considered
\cite{N2006,GN2006,GNO2006} a change of variables and use of
quasideterminant identities are necessary to achieve direct
verification of solutions.

\section{Preliminaries}\label{sec:prelim}

In this short section we recall some of the key elementary
properties of quasideterminants. The reader is referred to the
original papers \cite{GR,GGRL} for a more detailed and general
treatment.
\subsection{Quasideterminants}
An $n\times n$ matrix $A$ over a ring $\mathcal R$
(non-commutative, in general) has $n^2$ \emph{quasideterminants}
written as $|A|_{i,j}$ for $i,j=1,\dots, n$, which are also
elements of $\mathcal R$. They are defined recursively by
\begin{align*}
    |A|_{i,j}&=a_{i,j}-r_i^j(A^{i,j})^{-1}c_j^i,\quad A^{-1}=(|A|_{j,i}^{-1})_{i,j=1,\dots,n}.
\end{align*}
In the above $r_i^j$ represents the $i$th row of $A$ with the
$j$th element removed, $c_j^i$ the $j$th column with the $i$th
element removed and $A^{i,j}$ the submatrix obtained by removing
the $i$th row and the $j$th column from $A$. Quasideterminants can
also denoted as shown below by boxing the entry about which the
expansion is made
\[
|A|_{i,j}=\begin{vmatrix}
    A^{i,j}&c_j^i\\
    r_i^j&\fbox{$a_{i,j}$}
    \end{vmatrix}.
\]

The case $n=1$ is rather trivial; let $A=(a)$,  say,  and then
there is one quasideterminant $|A|_{1,1}=|\fbox{$a$}|=a$. For
$n=2$, let $A=\begin{pmatrix}   a&b\\c&d\end{pmatrix}$, then there
are four quasideterminants
\begin{align*}
|A|_{1,1}=\begin{vmatrix}
    \fbox{$a$}&b\\
    c&d
    \end{vmatrix}=a-bd^{-1}c,\quad
|A|_{1,2}=\begin{vmatrix}
    a&\fbox{$b$}\\
    c&d
    \end{vmatrix}=b-ac^{-1}d,\\
|A|_{2,1}=\begin{vmatrix}
    a&b\\
    \fbox{$c$}&d
    \end{vmatrix}=c-db^{-1}a,\quad
|A|_{2,2}=\begin{vmatrix}
    a&b\\
    c&\fbox{$d$}
    \end{vmatrix}=d-ca^{-1}b.
\end{align*}
From this we can obtain the matrix inverse,
\[
    A^{-1}=\begin{pmatrix}
      (a-bd^{-1}c)^{-1}&(c-db^{-1}a)^{-1}\\
      (b-ac^{-1}d)^{-1}&(d-ca^{-1}b)^{-1}
    \end{pmatrix},
\]
which is then used in the definition of the 9 quasideterminants of
a $3\times 3$ matrix. Note that if the entries in $A$ commmute,
the above becomes the familiar formula for the inverse of a
$2\times2$ matrix with entries expressed as ratios of
determinants. Indeed this is true for any size of square matrix;
if the entries in $A$ commute then
\begin{equation}\label{commute}
|A|_{i,j}=(-1)^{i+j}\frac{\det(A)}{\det(A^{i,j})}.
\end{equation}

In this paper we will consider only quasideterminants that are
expanded about a term in the last column, most usually the last
entry. For a block matrix
\begin{equation*}\label{block}
 \begin{pmatrix}
      A&B\\
      C&d
    \end{pmatrix}
\end{equation*}
where $d\in\mathcal R$, $A$ is a square matrix over $\mathcal R$
of arbitrary size and $B$, $C$ are column and row vectors over
$\mathcal R$ of compatible lengths, we have
\begin{equation*}
 \begin{vmatrix}
      A&B\\
      C&\fbox{$d$}
    \end{vmatrix}
    =d-CA^{-1} B.
\end{equation*}

\subsection{Invariance under row and column operations}
The quasideterminants of a matrix have invariance properties
similar to those of determinants under elementary row and column
operations applied to the matrix. Consider the following
quasideterminant of an $n\times n $ matrix;
\begin{equation}\label{invariance}
    \begin{vmatrix}
    \begin{pmatrix}
      E&0\\
      F&g
    \end{pmatrix}
    \begin{pmatrix}
      A&B\\
      C&d
    \end{pmatrix}
    \end{vmatrix}_{n,n}=
    \begin{vmatrix}
      EA&EB\\
      FA+gC&FB+gd
    \end{vmatrix}_{n,n}=g(d-CA^{-1}B)=g
    \begin{vmatrix}
      A&B\\
      C&d
    \end{vmatrix}_{n,n}.
\end{equation}
The above formula can be used to understand the effect on a
quasideterminant of certain elementary row operations involving
multiplication on the left. This formula excludes those operations
which add left-multiples of the row containing the expansion point
to other rows since there is no simple way to describe the effect
of these operations. For the allowed operations however, the
results can be easily described; left-multiplying the row
containing the expansion point by $g$ has the effect of
left-multiplying the quasideterminant by $g$ and all other
operations leave the quasideterminant unchanged. There is
analogous invariance under column operations involving
multiplication on the right.


\subsection{Noncommutative Jacobi Identity}
There is a quasideterminant version of Jacobi's identity for
determinants, called the noncommutative Sylvester's Theorem  by
Gelfand and Retakh \cite{GR}. The simplest version of this
identity is given by
\begin{equation}\label{nc syl}
    \begin{vmatrix}
      A&B&C\\
      D&f&g\\
      E&h&\fbox{$i$}
    \end{vmatrix}=
    \begin{vmatrix}
      A&C\\
      E&\fbox{$i$}
    \end{vmatrix}-
    \begin{vmatrix}
      A&B\\
      E&\fbox{$h$}
    \end{vmatrix}
    \begin{vmatrix}
      A&B\\
      D&\fbox{$f$}
    \end{vmatrix}^{-1}
    \begin{vmatrix}
      A&C\\
      D&\fbox{$g$}
    \end{vmatrix}.
\end{equation}

\section{Solutions obtained using Darboux transformations}\label{solutions}
\subsection{Darboux transformation}
Let $L$ be an operator covariant under the Darboux transformation
$G_\theta=\theta\partial_x\theta^{-1}=\partial_x-\theta_x\theta^{-1}$
where $\theta$ is an eigenfunction of $L$ (i.e.\ $L(\theta)=0$).
Let $\theta_i$, $i=1,\dots,n$ be a particular set of
eigenfunctions and introduce the notation
$\Theta=(\theta_1,\dots,\theta_n)$ and
$\widehat{\Theta}=(\theta_j^{(i-1)})_{i,j=1,\dots,n}$, the
$n\times n$ wronskian matrix of $\theta_1,\dots,\theta_n$, where
$^{(k)}$ denotes the $k$th $x$-derivative.

Let $\phi_{[1]}=\phi$ be a general eigenfunction of $L_{[1]}=L$
and $\theta_{[1]}=\theta_1$. Then
$\phi_{[2]}:=G_{\theta_{[1]}}(\phi_{[1]})$ and
$\theta_{[2]}=\phi_{[2]}|_{\phi\to\theta_2}$ are eigenfunctions
for $L_{[2]}=G_{\theta_{[1]}}L_{[1]}G_{\theta_{[1]}}^{-1}$. In
general, for $n\ge 1$ define the $n$th Darboux transform of $\phi$
by
\[
    \phi_{[n+1]}=\phi_{[n]}^{(1)}-\theta_{[n]}^{(1)}\theta_{[n]}^{-1}\phi_{[n]},
\]
in which
\[
    \theta_{[k]}=\left.\phi_{[k]}\right|_{\phi\to\theta_k}.
\]
It is known that \cite{GV,GGRL}
\begin{equation}\label{diter}
    \phi_{[n+1]}=
    \begin{vmatrix}
    \Theta&\phi\\
    \vdots&\vdots\\
    \Theta^{(n-1)}&\phi^{(n-1)}\\
    \Theta^{(n)}&\fbox{$\phi^{(n)}$}
    \end{vmatrix}.
\end{equation}

The Lax pair for the ncKP equation \eqref{nckp} is
\begin{align}
\label{lax1}    L&=\partial_x^2+v_x-\partial_y,\\
\label{lax2}
M&=4\partial_x^3+6v_x\partial_x+3v_{xx}+3v_y+\partial_t.
\end{align}
Both $L$ and $M$ are covariant with respect to the above Darboux
transformation. Moreover, it is straightforward to calculate that
the effect of the Darboux transformation
\[
\tilde L=G_\theta LG_{\theta}^{-1},\quad \tilde M=G_\theta
MG_{\theta}^{-1},
\]
is that $\tilde v=v+2\theta_x\theta^{-1}$. Thus after $n$ Darboux
transformations we obtain
\begin{equation}\label{nckp sol}
    v_{[n+1]}=v+2\sum_{i=1}^n\theta_{[i],x}\theta_{[i]}^{-1},
\end{equation}
which describes a class of solutions of ncKP. An analogous formula
is obtained using a noncommutative version of Gelfand-Dickey
theory in \cite{EGR1997}.
%
%
Further, it may be proved by induction from \eqref{diter}, making
use of an identity of the form \eqref{nc syl}, that
\begin{equation}\label{solu}
    \sum_{i=1}^n\theta_{[i],x}\theta_{[i]}^{-1}=-
    \begin{vmatrix}
    \Theta&0\\
    \vdots&\vdots\\
    \Theta^{(n-2)}&0\\
    \Theta^{(n-1)}&1\\
    \Theta^{(n)}&\fbox{0}
    \end{vmatrix}.
\end{equation}
Thus, using Darboux transformations, we have obtained a formula
for solutions $v_{[n+1]}$ of ncKP expressed in terms of a known
solution $v$ and a single wronskian-like quasideterminant,
\begin{equation}\label{nckp sol final}
    v_{[n+1]}=v-2
    \begin{vmatrix}
    \Theta&0\\
    \vdots&\vdots\\
    \Theta^{(n-2)}&0\\
    \Theta^{(n-1)}&1\\
    \Theta^{(n)}&\fbox{0}
    \end{vmatrix}.
\end{equation}
%
%
\subsection{Binary Darboux transformation}
To define a binary Darboux transformation one needs to consider
the adjoint Lax pair. The notion of adjoint is easily extended
from the familiar matrix situation to any ring $\mathcal R$ (see
\cite{M1997}); suppose that for each $a\in\mathcal R$, there
exists $a^\dagger\in\mathcal R$, for any derivative $\partial$
acting on $\mathcal R$, $\partial^\dagger=-\partial$ and for any
product $AB$ of elements of, or operators on $\mathcal R$,
$(AB)^\dagger=B^\dagger A^\dagger$. Accordingly, the adjoint Lax
pair is
\begin{align}\label{ad lax}
    L^\dagger&=\partial_x^2+v^\dagger_x+\partial_y,\\
    M^\dagger&=-4\partial_x^3-6v^\dagger_x\partial_x-3v^\dagger_{xx}+3v^\dagger_y-\partial_t.
\end{align}

Following the standard construction of a binary Darboux
transformation (see \cite{MS,OS}) one introduces a potential
$\Omega(\phi,\psi)$ satisfying
\begin{equation}\label{Omega}
\Omega(\phi,\psi)_x=\psi^\dagger\phi,\quad
\Omega(\phi,\psi)_y=\psi^\dagger\phi_x-\psi^\dagger_x\phi,\quad
\Omega(\phi,\psi)_t=-4(\psi^\dagger\phi_{xx}-\psi^\dagger_x\phi_x+\psi^\dagger_{xx}\phi)-6\psi^\dagger
v_x\phi.
\end{equation}
The parts of this definition are compatible when
$L(\phi)=M(\phi)=0$ and $L^\dagger(\psi)=M^\dagger(\psi)=0$. More
generally, we can define $\Omega(\Phi,\Psi)$ for any row vectors
$\Phi$ and $\Psi$ such that $L(\Phi)=M(\Phi)=0$ and
$L^\dagger(\Psi)=M^\dagger(\Psi)=0$. If $\Phi$ is an $n$-vector
and $\Psi$ is an $m$-vector then $\Omega(\Phi,\Psi)$ is an
$m\times n$ matrix. The adjoint of a $p\times q$ matrix
$A=(a_{i,j})$ over $\mathcal R$ has an obvious meaning. It is the
$q\times p$ matrix $A^\dagger=(a^\dagger_{j,i})$.

A binary Darboux transformation is then defined by
\[
    \phi_{[n+1]}=\phi_{[n]}-\theta_{[n]}\Omega(\theta_{[n]},\rho_{[n]})^{-1}\Omega(\phi_{[n]},\rho_{[n]})
\]
and
\[
    \psi_{[n+1]}=\psi_{[n]}-\rho_{[n]}\Omega(\theta_{[n]},\rho_{[n]})^{-\dagger}\Omega(\theta_{[n]},\psi_{[n]})^\dagger,
\]
where
\[
    \theta_{[n]}=\left.\phi_{[n]}\right|_{\phi\to\theta_n},\quad
    \rho_{[n]}=\left.\psi_{[n]}\right|_{\psi\to\rho_n}.
\]
Using the notation $\Theta=(\theta_1,\dots,\theta_n)$ and
$\Rho=(\rho_1,\dots,\rho_n)$ it is easy to prove by induction that
for $n\ge 1$,
\begin{equation}\label{B phi}
    \phi_{[n+1]}=
    \begin{vmatrix}
    \Omega(\Theta,\Rho)&\Omega(\phi,\Rho)\\
    \Theta&\fbox{$\phi$}
    \end{vmatrix},
\end{equation}
\begin{equation}\label{B psi}
    \psi_{[n+1]}=
    \begin{vmatrix}
    \Omega(\Theta,\Rho)^\dagger&\Omega(\Theta,\psi)^\dagger\\
    \Rho&\fbox{$\psi$}
    \end{vmatrix},
\end{equation}
and
\begin{equation}\label{Omega again}
   \Omega(\phi_{[n+1]},\psi_{[n+1]})=
    \begin{vmatrix}
    \Omega(\Theta,\Rho)&\Omega(\phi,\Rho)\\
    \Omega(\Theta,\psi)&\fbox{$\Omega(\phi,\psi)$}
    \end{vmatrix}.
\end{equation}
As for the effect of these transformation on the Lax pair, a
transformation by $\theta,\rho$ gives new coefficients defined in
terms of
\[
    \hat v=v+2\theta\Omega(\theta,\rho)^{-1}\rho^\dagger.
\]
Thus after $n$ binary Darboux transformations we obtain
\begin{equation}
  v_{[n+1]}=v+2\sum_{k=1}^n\theta_{[k]}\Omega(\theta_{[k]},\rho_{[k]})^{-1}\rho_{[k]}^\dagger,
\end{equation}
and this may be reexpressed in terms of a single quasideterminant
as
\begin{equation}\label{bin sol}
  v_{[n+1]}=v-2
    \begin{vmatrix}
    \Omega(\Theta,\Rho)&\Rho^\dagger\\
    \Theta&\fbox{$0$}
    \end{vmatrix}.
\end{equation}
In this way we have obtained a second expression for solutions of
the ncKP equation, this time in terms of grammian-type
quasideterminants.

In the following sections, we will show that these two
quasideterminant solutions (wronskian-type and grammian-type) of
the ncKP equation may also be verified by direct calculation in
the spirit of Hirota's direct method.

\section{Derivatives of a quasideterminant}\label{derivatives}
We can derive a rather appealing formula for derivatives of a
quasideterminant which resembles the formula for derivatives of a
normal determinant. Consider the derivative
\begin{equation}\label{blockdash}
 \begin{vmatrix}
      A&B\\
      C&\fbox{$d$}
    \end{vmatrix}'
    =d'-C' A^{-1} B +CA^{-1} A' A^{-1} B-CA^{-1} B'
\end{equation}
where $A$ is an $n \times n$ matrix, $C$ is a row vector and $B$ a
column vector. If the matrix $A$ has the grammian-like property
that its derivative is a scalar product
\[
    A'=\sum_{i=1}^kE_iF_i,
\]
where $E_i$ ($F_i$) are column (row) vectors of appropriate
length, then the third term on the RHS of (\ref{blockdash}) can be
factorised as a product of quasideterminants, i.e.
\begin{align}
    \begin{vmatrix}
    A&B\\
    C&\fbox{$d$}
    \end{vmatrix}'&=
    d'-C'A^{-1}B+\sum_{i=1}^k(CA^{-1}E_i)(F_i A^{-1}B)- CA^{-1} B'\nonumber\\
\label{gram deriv}
  &=
   \begin{vmatrix}
    A&B\\
    C'&\fbox{$d'$}
    \end{vmatrix}+\begin{vmatrix}
    A&B'\\
    C&\fbox{$ 0$}
    \end{vmatrix}+\sum_{i=1}^k\begin{vmatrix}
    A&E_i\\
    C&\fbox{$0$}
    \end{vmatrix}\begin{vmatrix}
    A&B\\
    F_i&\fbox{$0$}
    \end{vmatrix}.
\end{align}
Even if the matrix $A$ does not have this grammian-like structure
then the third term on the RHS of (\ref{blockdash}) can still be
factorized as a product by inserting the $n \times n$ identity
matrix expressed in the form
\[
I=\sum_{k=1}^ne_ke_k^t,
\]
where $e_k$ is the $n$-vector $(\delta_{ik})$ (i.e.\ a column
vector with 1 in the $k$th row and 0 elsewhere). Let $Z^k$ denote
the $k$th row and $Z_k$ the $k$th column of a matrix $Z$. In this
way we have
\begin{align*}
    \begin{vmatrix}
    A&B\\
    C&\fbox{$d$}
    \end{vmatrix}'&=
    d'-C'A^{-1}B+\sum_{k=1}^n(CA^{-1}e_k)(e_k^tA'A^{-1}B)-\sum_{k=1}^n (CA^{-1}e_k) e_k^t B'.
\end{align*}
Note here that we have also introduced this form of the identity
into the last term on the RHS. This gives
\begin{equation}\label{der}
    \begin{vmatrix}
    A&B\\
    C&\fbox{$d$}
    \end{vmatrix}'=
    \begin{vmatrix}
    A&B\\
    C'&\fbox{$d'$}
    \end{vmatrix}
    +\sum_{k=1}^n
    \begin{vmatrix}
    A&e_k\\
    C&\fbox{$0$}
    \end{vmatrix}
     \begin{vmatrix}
    A&B\\
    (A^k)'&\fbox{$(B^k)'$}
    \end{vmatrix}
.
\end{equation}
In a similar way, by inserting the identity matrix in a different
position we have a column version of the derivative formula
\begin{align}
\label{col diff} \begin{vmatrix}
    A&B\\
    C&\fbox{$d$}
    \end{vmatrix}' &=
    \begin{vmatrix}
    A&B'\\
    C&\fbox{$d'$}
    \end{vmatrix}
    +\sum_{k=1}^n
    \begin{vmatrix}
    A&(A_k)'\\
    C&\fbox{$(C_k)'$}
    \end{vmatrix}
    \begin{vmatrix}
    A&B\\
    e_k^t&\fbox{$0$}
    \end{vmatrix}.
\end{align}

\subsection{Derivatives of quasiwronskians}

In this section we will calculate derivatives of a
quasideterminant of the form
\begin{equation}
   Q(i,j)=
    \begin{vmatrix}
    \widehat{\Theta}&e_{n-j}\\
\Theta^{(n+i)}&\fbox{0}
    \end{vmatrix},
\end{equation}
where, as above,
$\widehat{\Theta}=(\theta_j^{(i-1)})_{i,j=1,\dots,n}$ is the
$n\times n$ wronskian matrix of $\theta_1,\dots,\theta_n$. In this
definition, $i$ and $j$ are allowed to take any integer values
subject to the convention that if $n-j$ lies outside the range
$1,2,\dots,n$, then $e_{n-j}=0$ and so $Q(i,j)=0$. There is an
important special case; when $n+i=n-j-1\in[0,n-1]$, (i.e.\
$i+j+1=0$ and $-n\le i<0$) we have
\[
   Q(i,j)=
    \begin{vmatrix}
    \Theta&0\\
    \vdots&\vdots\\
    \Theta^{(n+i)}&1\\
    \vdots&\vdots\\
    \Theta^{(n-1)}&0\\
    \Theta^{(n+i)}&\fbox{0}
    \end{vmatrix}=
    \begin{vmatrix}
    \Theta&0\\
    \vdots&\vdots\\
    \Theta^{(n+i)}&1\\
    \vdots&\vdots\\
    \Theta^{(n-1)}&0\\
    0&\fbox{-1}
    \end{vmatrix}=-1,
\]
using the definition of quasideterminants and the invariance
properties described in \eqref{invariance}. Using the same
argument for $n+i\in[0,n-1]$ but $n+i\ne n-j-1$ we see that
$Q(i,j)=0$. Assuming $n$ is arbitrarily large we may summarise
these properties of $Q(i,j)$ as
\begin{equation}\label{Q prop}
    Q(i,j)=\begin{cases}
      -1&i+j+1=0\\
      0&(i<0\text{ or }j<0)\text{ and }i+j+1\ne0.
    \end{cases}
\end{equation}
Readers familiar with symmetric functions will recognise this
property as analogous to a property of a hook Schur function
$s_{(i|j)}$ (see \cite[p47, Ex. 9]{Macdonald}).

We shall call this type of quasideterminant a
\emph{quasiwronskian}. In the last section (see \eqref{nckp sol})
we showed by means of Darboux transformations that if $v_0$ is any
given solution of ncKP and $\Theta$ an $n$-row vector of
eigenfunctions of $L$ and $M$ given by \eqref{lax1}--\eqref{lax2}
then
\begin{equation}
    v=v_0-2Q(0,0),
\end{equation}
also satisfies ncKP. For simplicity we will choose the vacuum
solution $v_0=0$ but this choice of vacuum is not essential to
what follows; the direct verification can be made for arbitrary
vacuum but the formulae are rather more complicated.

If we relabel and rescale the independent variables so that
$x_1=x$, $x_2=y$, $x_3=-4t$, $\Theta$ satisfies the linear
equations
\begin{align}\label{disper}
\begin{split}
\Theta_{x_2}&=\Theta_{xx},\\
\Theta_{x_3}&=\Theta_{xxx}.
\end{split}
\end{align}
We may also allow $\Theta$ to depend on higher variables $x_k$ and
impose the natural dependence
$\Theta_{x_k}=\Theta_{\underbrace{x\cdots x}_{\text{$k$
copies}}}$.

Now for any $m$, using the linear equations for $\Theta$, we have
\begin{align}
\frac{\partial}{\partial x_m}  Q(i,j)&=
    \begin{vmatrix}
    \widehat{\Theta} &e_{n-j}\\
    \;\Theta^{(n+i+m)}&\fbox{0}\;
    \end{vmatrix}
    +\sum_{k=1}^n \begin{vmatrix}
    \widehat{\Theta} &e_{k}\\
    \;\Theta^{(n+i)}&\fbox{0}\;
    \end{vmatrix}\begin{vmatrix}
    \widehat{\Theta} &e_{n-j}\\
    \;\Theta^{(k-1+m)}&\fbox{0}\;
    \end{vmatrix} \nonumber\\
    &=Q(i+m,j)+ \sum_{k=0}^{n-1}  Q(i,k)Q(m-1-k,j).\label{argh}
\end{align}
Using the conditions \eqref{Q prop} the above simplifies
considerably and we obtain
\begin{equation}\label{quasiwr}
\frac{\partial}{\partial x_m} Q(i,j)= Q(i+m,j)-Q(i,j+m) +
\sum_{k=0}^{m-1}  Q(i,k)\; Q(m-k-1,j).
\end{equation}
In particular
\begin{align*}
 \frac{\partial}{\partial x}  Q(i,j)
 &=Q(i+1,j)-Q(i,j+1)+ Q(i,0) \;Q(0,j),\\
\frac{\partial}{\partial x_2} Q(i,j)
&= Q(i+2,j)-Q(i,j+2) +Q(i,1)\; Q(0,j)+Q(i,0)\; Q(1,j),\\
\frac{\partial}{\partial x_3} Q(i,j) &= Q(i+3,j)-Q(i,j+3)
+Q(i,2)\; Q(0,j)+Q(i,1)\; Q(1,j)+Q(i,0)\; Q(2,j).
\end{align*}

Note that these simplified formulae \eqref{quasiwr} are only valid
for sufficiently large $n$. For smaller $n$ we should use
\eqref{argh} directly.

\subsection{Derivatives of quasigrammians}
Let us define
\[
R(i,j)=(-1)^{j}\begin{vmatrix}
    \Omega(\Theta,\Rho)&\Rho^{\dagger (j)}\\
    \Theta^{(i)}&\fbox{$0$}
    \end{vmatrix},
\]
and call this type of quasideterminant a \emph{quasigrammian}. As
we have seen in \eqref{bin sol}, solutions obtained by binary
Darboux transformation are of the form $v=v_0-2R(0,0)$. As we did
in the case of the quasiwronskian type of solutions we choose
$v_0=0$ for simplicity. Hence $\Theta$ satisfies the same linear
equations as before and $\Rho$, the vector of adjoint
eigenfunctions, satisfies
\[
\Rho_{x_2}=- \Rho_{xx},\qquad \Rho_{x_3}= \Rho_{xxx}.
\]
Note that choice of the trivial vacuum is inessential and direct
verification can be completed for arbitrary vacuum.

Using \eqref{gram deriv}, derivatives with respect to the $x_m$
can be calculated;
\begin{align*}
\partial_{x_m} R(i,j)&=
    (-1)^j \begin{vmatrix}
    \Omega&\Rho^{\dagger (j)}\\
    \Theta^{(i+m)}&\fbox{$0$}
    \end{vmatrix}
    +(-1)^{m+j-1}\begin{vmatrix}
    \Omega&\Rho^{\dagger (j+m)}\\
    \Theta^{(i)}&\fbox{$0$}
    \end{vmatrix}\\&\\
    &\qquad\qquad\qquad\qquad\qquad\qquad+
    \sum^{m-1}_{k=0}\begin{vmatrix}
    (-1)^{j+k}
    \Omega&\Rho^{\dagger (k)}\\
    \Theta^{(i)}&\fbox{$0$}
    \end{vmatrix}
    \begin{vmatrix}
    \Omega&\Rho^{\dagger (j)}\\
    \Theta^{(m-1-k)}&\fbox{$0$}
    \end{vmatrix}\\
    &=R(i+m,j)-R(i,j+m)+\sum_{k=0}^{m-1} R(i,k) R(m-k-1,j).
\end{align*}
Notice here that this final form for a derivative of a
quasigrammian corresponds precisely with the formula for the
quasiwronskian (see (\ref{quasiwr})). Thus calculations in the
subsequent sections carried out for the quasiwronskian solutions
will be equally valid for the quasigrammian solutions.

\subsection{The commutative case}

In order to better understand the derivative formulae we obtained
above, we will assume that all quantities commute and hence reduce
to the familiar case of the commutative KP equation. Using
\eqref{commute}, we have
\begin{equation*}
Q(0,0)=-\frac{
\begin{vmatrix}
    \Theta \\
    \vdots\\
    \Theta^{(n-2)}\\
    \Theta^{(n)}
    \end{vmatrix}
    }{\begin{vmatrix}
    \widehat{\Theta}
    \end{vmatrix}},
\qquad\qquad R(0,0)=\frac{\begin{vmatrix}
    \Omega&\Rho\\
    \Theta&\fbox{$0$}
    \end{vmatrix}}{
    \begin{vmatrix}
    \Omega
    \end{vmatrix}}.
    \end{equation*}
It is then simple to show that
$u=-2Q(0,0)_x=2(\log|\widehat{\Theta}|)_{xx}$ and
$u=-2R(0,0)_x=2(\log|\Omega|)_{xx}$ which are the well-known
solutions of the standard KP solution in wronskian and grammian
form respectively.

\section{The direct approach}
Returning to the noncommutative case, we will show directly that
\begin{equation}\label{sol}
v=-2 Q(0,0), \qquad \text{or}\qquad v=-2 R(0,0),
\end{equation}
are solutions of the ncKP equation. To carry out this direct
verification we first calculate the derivatives of $v$
\begin{align*}
v_x&=-2 Q(0,0)_x=-2[Q(1,0)-Q(0,1)+  Q( 0,0 ) Q(0 ,0 )]\\
v_y&=-2 Q(0,0)_y=-2[Q(2,0)-Q(0,2)+  Q( 0, 0) Q(1 ,0 )+ Q( 0, 1) Q(0 ,0 )]\\
v_t&= -2 Q(0,0)_{t}=8 [Q(3,0)-Q(0,3)+  Q( 0, 0) Q(2 ,0 )+ Q( 0, 1)
Q(1 ,0 )+Q( 0, 2) Q(0 ,0 )]\\
v_{xx}&=-2 [Q(0,2) - 2 Q(1,1) + Q(2,0) - 2 Q(0,0) Q(0,1) +
Q(0,0)Q(1,0) - Q(0,1) Q(0,0)\\&\qquad+2 Q(1,0)Q(0,0) + 2 Q(0,0) Q(0,0) Q(0,0)]\\
v_{yy}&=-2[Q(0, 4) - 2 Q(2, 2) + Q(4, 0)\\
&\qquad+  Q(0, 0) Q(3, 0)+  Q(0, 1) Q(2, 0)-  Q(0, 2) Q(1, 0)-
    Q(0, 3) Q(0, 0)\\
    &\qquad-2  Q(0, 0) Q(1, 2)  -
  2  Q(0, 1) Q(0, 2)
    + 2  Q(2, 0) Q(1, 0) +
  2  Q(2, 1) Q(0, 0)\\ &\qquad+ 2  Q(0, 0) Q(1, 0) Q(1, 0) +
  2  Q(0, 0) Q(1, 1) Q(0, 0)\\ &\qquad+ 2  Q(0, 1)Q(0, 0) Q(1, 0) +
  2  Q(0, 1) Q(0, 1) Q(0, 0)]\\
v_{xt}&= 8[Q(0, 4) - Q(1, 3) - Q(3, 1) + Q(4, 0)\\
&\qquad+  Q(0, 0) Q(3, 0)  -  Q(0, 3) Q(0, 0)-  Q(0, 0) Q(2, 1)-
Q(0, 2) Q(0, 1)-
   Q(0, 1) Q(1, 1)\\
   &\qquad- Q(0, 0) Q(0, 3)    +
   Q(1, 0) Q(2, 0) +  Q(1, 1) Q(1, 0)) +  Q(1, 2) Q(0, 0)) +
   Q(3, 0) Q(0, 0) \\
   &\qquad+  Q(0, 0) Q(0, 0) Q(2, 0) +
   Q(0, 0)Q(0, 1) Q(1, 0) +  Q(0, 0) Q(0, 2) Q(0, 0)\\
    &\qquad+
   Q(0, 0) Q(2, 0) Q(0, 0) +  Q(0, 1) Q(1, 0) Q(0, 0) +
   Q(0, 2) Q(0, 0)Q(0, 0)]
\end{align*}
and $v_{xxxx}$, which is straightforward but tedious to work out.
This sort of calculation can be readily carried out using any
computer algebra package that understands, or can be made to
understand, noncommutative multiplication. Substituting these into
the ncKP equation \eqref{nckp} all terms exactly cancel and the
solution is verified. As remarked above, the derivative formulae
are the same whether we use the quasiwronskian or the
quasigrammian formulation and so the above calculation
simultaneously verifies both types of solution.

\section{Comparison with the bilinear approach}
The direct approach to the determinantal solutions of the
commutative KP equation is well known and can be found in many
places in the literature (see \cite{FN,Hirota book} for example).
Here we will compare it to the alternative direct approach we
studied above. In Hirota's direct method, one first makes the
change of variables
\[
u=2 (\log \tau)_{xx}
\]
and then rewrites \eqref{kp} in bilinear form using Hirota
derivatives
\begin{equation}\label{Hiroteq}
(D_{xt}+D_{xxxx}+3 D_{yy})\tau \cdot \tau=0,
\end{equation}
where \cite{Hirota book}
\[
    D_x^mD_y^n\tau\cdot\tau:=\left.\frac{\partial^m\partial^n}{\partial a^m\partial b^m}
    \biggl(\tau(x+a,y+b)\tau(x-a,y-b)\biggr)\right|_{a,b\to0}.
\]
The next step is to take a possible solution such as a wronskian
or grammian determinant and calculate the derivatives with respect
to $x$, $y$ and $t$.  So, for example, for a wronskian solution we
would take, (see \cite{FN} for an explanation of the notation)
\begin{equation*}\label{bilin}
 \tau=\begin{vmatrix}
    \widehat{\Theta}
    \end{vmatrix},
\end{equation*}
and calculate the derivatives
\begin{align*}
 \tau_x&=(0,\cdots,n-2,n)=\tau_{(1)},\\
    \tau_{xx}&=(0,\cdots,n-2,n+1)+(0,\cdots,n-3,n-1,n)=\tau_{(2)}+\tau_{(1^2)},\\
    \tau_y&=\tau_{(2)}-\tau_{(1^2)},
\end{align*}
and so on. Here we use a shorthand partition notation which
denotes the extra derivatives added to each row in the wronksian
$\tau$.

Substituting these into the left hand side of \eqref{Hiroteq} we
obtain a constant multiple of
\begin{equation}\label{bb}
\tau_{(2^2)}\tau-\tau_{(21)}\tau_{(1)}+\tau_{(2)}\tau_{(1^2)}.
\end{equation}
While it is initially not obvious that this expression is
identically zero, using the Laplace expansion of a $2n\times 2n$
determinant, one verifies that \eqref{bb} is indeed zero and the
verification is complete.

For the noncommutative case the approach is quite similar,
however, curiously some of the steps taken in the bilinear
approach are not needed. First, we do not need a Cole-Hopf style
change of variables, since the solution is expressed directly as a
quasiwronskian. Second, once we have substituted the derivatives
into the nonlinear equation the resulting expression immediately
vanishes without the need to consider any quasideterminant
identities.

%
The fact that no identities are needed is rather unexpected but a
closer examination of the derivatives of $Q(0,0)$ in the
commutative case is illuminating. When all quantities commute we
may use \eqref{commute} to obtain
\[
    Q(i,j)=(-1)^{j-1}\frac{\tau_{(i+1,1^j)}}{\tau},
\]
and in particular $Q(0,0)=-\tau_{(1)}/\tau=-\tau_{x}/\tau$.
Calculating the $t$ derivative of each side of this gives
\begin{align*}
    -\tfrac14Q(0,0)_{t}&=Q(3,0)-Q(0,3)+Q(0,0) Q(2 ,0 )+ Q( 0, 1)Q(1 ,0 )+Q( 0, 2) Q(0 ,0 )\\
    &=-\frac{\tau_{(1^4)}+\tau_{(4)}}\tau+\frac{\tau_{(1)}\tau_{(3)}
    -\tau_{(1^2)}\tau_{(2)}+\tau_{(1^3)}\tau_{(1)}}{\tau^2},
\end{align*}
whereas
\begin{align*}
    -\frac14\left(\frac{-\tau_x}{\tau}\right)_{t}&=
    -\frac14\left(-\frac{\tau_{xt}}{\tau}+\frac{\tau_{x}\tau_{t}}{\tau^2}\right)\\
    &=-\frac{\tau_{(1^4)}-\tau_{(2^2)}+\tau_{(4)}}\tau
    +\frac{\tau_{(1)}(\tau_{(3)}-\tau_{(21)}+\tau_{(1^3)})}{\tau^2}.
\end{align*}
Note that the term $\tau_{(2^2)}$ cannot come from $Q(i,j)$ for
any $i,j$ and that the two expression for the derivatives only
agree when one makes use of the identity
$\tau_{(2^2)}\tau-\tau_{(21)}\tau_{(1)}+\tau_{(2)}\tau_{(1^2)}=0$.
So it seems that, in some sense, the identity used by hand in
verifying solutions in the bilinear approach, is used
automatically as derivatives are calculated in the
quasideterminant approach.

\section{Conclusions}
In this paper we considered two types of quasideterminant
solutions of the noncommutative KP equation. As well as showing
how they may be constructed by Darboux transformations, they turn
out to be ideal for direct verification of the solution and play
the same role that the $\tau$-function does in the commutative
case.

There are some interesting features to the direct approach using
quasideterminants. First, it illustrates that a bilinear form is
not needed, and indeed we believe that it does not exist, in the
noncommutative case. The second rather surprising feature is that,
unlike the commutative case, no identity is needed to complete the
verification.

Noncommutative versions of other integrable equations we have
studied, a noncommutative Hirota-Miwa equation
\cite{N2006,GNO2006} and modified KP equation \cite{GN2006}, also
have quasideterminant solutions. However, in these cases, direct
verification does require the use of quasideterminant identities
of the form \eqref{nc syl} and so it seems that the ncKP equation
may be exceptional in this respect.


\begin{thebibliography}{99}
\bibitem{HT1}
M. Hamanaka and K. Toda, \textit{Phys. Lett. A} {\bf 316} 77--83
(2003).


\bibitem{P}
L. D. Paniak, ``Exact Noncommutative KP and KdV Multi-solitons''
hep-th/0105185 (2001).


\bibitem{S}
M. Sakakibara, \textit{J. Phys A: Math. Gen.} {\bf 37} L599--L604
(2004).


\bibitem{WW1}
N. Wang and M. Wadati, \textit{J. Phys. Soc. Jap.} {\bf 72}
1366--1373 (2003).

\bibitem{WW2}
N. Wang and M. Wadati, \textit{J. Phys. Soc. Jap.} {\bf 72}
p1881-1888 (2003).

\bibitem{WW3}
N. Wang and M. Wadati, \textit{J. Phys. Soc. Jap.} {\bf 73}
p1689-1698 (2003).

\bibitem{H}
M. Hamanaka, ``Noncommutative solitons and D-branes'' PhD thesis
and hep-th/0303256 (2003).


\bibitem{K}
B. A. Kupershmidt, \emph{KP or mKP. Noncommutative mathematics of
Lagrangian, Hamiltonian, and integrable systems}. Mathematical
Surveys and Monographs, 78. Amer. Maths. Soc. (2000).


\bibitem{EGR1997}
P. Etingof, I. Gelfand and V. Retakh, \textit{Maths. Res. Lett.}
{\bf 4} 413--425 (1997).


\bibitem{GR}
I. Gelfand and V. Retakh, \textit{Funct. Anal. Appl.} {\bf 25}
91-102 (1991).


\bibitem{GGRL}
I. Gelfand, S. Gelfand, V. Retakh and R. L. Wilson, \textit{Adv.
in Math.} {\bf 193} 56-141 (2005)


\bibitem{H2006}
M. Hamanaka, ``Notes on Exact Multi-Soliton Solutions of
Noncommutative Integrable Hierarchies'' hep-th/0610006 (2006).


\bibitem{GV}
V. M. Goncharenko and A. P. Veselov, \textit{J. Phys. A: Math.
Gen.} {\bf 31} 5315-­5326 (1998).


\bibitem{FN}
N. C. Freeman and J. J. C. Nimmo,  \textit{Phys. Lett. A}
\textbf{95} 1--3 (1983).


\bibitem{Hirota book}
R. Hirota, \emph{The Direct Method in Soliton Theory}, Cambridge
University Press, Cambridge (2004).


\bibitem{N2006}
J. J. C. Nimmo, \textit{J. Phys. A: Math. Gen.} \textbf{39}
5053-5065 (2006).


\bibitem{GN2006}
C. R. Gilson and J. J. C. Nimmo, ``On quasideterminant solutions
of a noncommutative modified KP equation'', in preparation.


\bibitem{GNO2006}
C. R. Gilson, J. J. C. Nimmo and Y. Ohta, ``Quasideterminant
solutions of a non-Abelian Hirota-Miwa equation'', in preparation.





\bibitem{M1997}
V. B. Matveev,  ``Darboux transformations in differential rings
and functional-difference equations"  A. Karman (ed.) ,
\textit{Proc. CRM Workshop Bispectral Problems} (Montreal March,
1997) , Amer. Math. Soc. (1998).


\bibitem{MS}
V. B. Matveev and M. A. Salle, \textit{Darboux transformations and
solitons}, Springer Series in Nonlinear Dynamics. Springer-Verlag,
Berlin, (1991).


\bibitem{OS}
W. Oevel and W. Schief,
 ``Darboux theorems and the KP hierarchy" in \textit{Applications of analytic
 and geometric methods to nonlinear differential equations} (Exeter, 1992),
 193--206, NATO Adv. Sci. Inst. Ser. C Math. Phys. Sci., 413,
 Kluwer Acad. Publ., Dordrecht, (1993).


\bibitem{Macdonald}
I. D. Macdonald, \textit{Symmetric functions and Hall polynomials}
(2nd Edition) Oxford University Press, Oxford (1995).







\end{thebibliography}
\end{document}